\documentclass[fleqn,usenatbib]{mnras}

\usepackage{mathptmx}
\usepackage[T1]{fontenc}
\usepackage{ae,aecompl}
\usepackage{color}
\usepackage{multirow}

\usepackage{graphicx}	
\usepackage{amsmath}	
\usepackage{amssymb}	
\usepackage{subfigure}

\newcommand{\imp}{\ensuremath{\textrm{imp}}}

\title[Production and fate of small particles due to  Aegaeon ]
{ Production and fate of the G~ring arc particles due to  Aegaeon~(Saturn~LIII)}

\author[Madeira {\sl et al.}]{
Gustavo Madeira\thanks{E-mail: gtvmadeira@feg.unesp.br},
R.~Sfair, D.C.~Mour\~ao and S.M.~Giuliatti Winter\\
Univ. Estadual Paulista -UNESP,  
Grupo de Din\^amica Orbital e Planetologia,
Guaratinguet\' a, CEP 12516-410, Brazil
}

\date{}

\begin{document}
\label{firstpage}
\pagerange{\pageref{firstpage}--\pageref{lastpage}}
\maketitle

\begin{abstract}
The G~ring arc hosts the smallest satellite of Saturn, Aegaeon, observed with a set of images sent by Cassini spacecraft. 
Along with Aegaeon, the arc particles are trapped in a 7:6 corotation eccentric resonance with the satellite Mimas. Due to this resonance, both Aegaeon and the arc material are confined to within sixty degrees of corotating longitudes. The arc particles are dust grains which can have their orbital motions severely disturbed by the solar radiation force. Our numerical simulations showed that Aegaeon is responsible for depleting the arc dust population by removing them through collisions. The solar radiation force hastens these collisions by removing most of the 10~$\mu$m sized grains in less than 40~years. Some debris released from Aegaeon's surface by meteoroid impacts can populate the arc. However, it would take 30,000 years for Aegaeon to supply the observed amount of arc material, and so it is unlikely that Aegaeon alone is the source of dust in the arc.
\end{abstract}

\begin{keywords}
planets and satellites: rings
\end{keywords}



\section{Introduction}
Before the discovery of the small satellite Aegaeon, a bright arc close to the inner edge of Saturn's G~ring was imaged by the cameras onboard the Cassini spacecraft. Located at about 167500~km from Saturn's centre, this arc extends over $\sim 60^{\circ}$ in longitude and has a radial  width of  250~km, while the rest of the G~ring is 6000~km wide \citep{He07}. Cassini  data showed that most of the arc is populated by $\mu$m sized dust grains, although larger bodies (cm to meters in size) can also be present. \cite{He07} argued that a decrease in the flux of energetic electrons, observed by the Cassini instruments, could be caused by a population of cm-m sized  bodies. They proposed that these large bodies could be the source of the arc and also the  G~ring.

The mean motion of the arc is close to the 7:6 corotation eccentric resonance (CER) with  Mimas \citep{He07}. The resonant argument  $\phi = 7 \lambda_{\rm M} - 6 \lambda - \varpi_{\rm M}$ is equal to 
$180^{\circ}$, where $\lambda_{\rm M}$ and  $\lambda$ are the mean longitudes of Mimas and the particle, respectively, and $\varpi_{\rm M}$ is the longitude of Mimas' pericenter. \cite{He07} numerically simulated a sample of particles initially located in this arc and verified that they stay confined for at least 80~years.

Several Cassini images taken between 2007 and 2009 showed  a small satellite, named Aegaeon, embedded in the G~ring arc. With a diameter about 500~m, Aegaeon is trapped in the same 7:6 corotation eccentric resonance with Mimas with a libration amplitude of about $10^{\circ}$ \citep{He10}.

In this work we analyze the influence of Aegaeon on the small particles located in the G~ring arc, after their possible ejection from the surface of Aegaeon, as well as their orbital evolution due to gravitational and dissipative forces. In section~\ref{arc} we analyze the orbital evolution of a set of $\mu m$ sized particles under the effects of the solar radiation force and the gravitational effects of the planet and the saturnian satellites, Mimas, Tethys and Aegaeon. In section~\ref{ejected} we analyze the time evolution of those dust particles ejected from the surface of the small satellite. Section~\ref{mass} compares these estimates of dust lifetimes with the production rate due to impacts onto Aegaeon's surface in order to analyze the role of the satellite on the maintenance of the arc population. Our results are discussed in the last section.

\section {Orbital evolution of the G~ring arc particles} \label{arc}

First of all we analyze the gravitational effects of Mimas on the particles located in the G~ring arc. 
The dynamical system is formed by Saturn, including the gravitational coefficients $J_2$, $J_4$ and 
$J_6$, and  the  satellites Mimas, Aegaeon and Tethys. 
The numerical simulations were performed using the Mercury integrator package \citep{Ch99} with 
the Burlish St\"oer algorithm. We also used the algorithm described in \cite{Re06} to convert the state 
vector into the geometric orbital elements that account for the orbital precession caused by the gravity coefficients of Saturn. 
Table~\ref{sats} shows the initial osculating elements (2454700.5 JD), mass and density of the satellites, while Table~\ref{saturn} presents the parameters of Saturn: radius (in km), mass (in kg) \citep{Th13},  $J_2$, $J_4$ and $J_6$ [consistent with \cite{He10}].

\begin{table}
\caption{Mass ($m$), density ($d$) and osculating elements of the satellites (2454700.5 JD): $a$ is the semi-major axis, $e$ is the eccentricity, $I$ is the inclination, and the angles $\varpi$, $\Omega$ and $\lambda$ are the argument of pericentre, longitude of node and mean anomaly, respectively. These values were extract from JPL-Horizons System.}
\centering
\label{sats}
\begin{tabular}{llll}
\hline \hline 
& Aegaeon & Mimas & Tethys \\ \hline
$a$ ($\times 10^5$km) & $1.6803396819$ & $1.8600466879$ & $2.9497426488$ \\
$e$ ($\times 10^{-2}$) & $0.3121285727$ & $1.7245219209$ & $0.0828116581$\\
$I$($^{\circ}$) & $0.0014761087$ & $1.5687571620$ & $1.0915162973$\\
$\varpi$ ($^{\circ}$) & $145.76280592$ & $163.18023984$ & $15.906312542$ \\
$\Omega$ ($^{\circ}$) & $233.02211168$ & $259.15258436$ & $355.42083194$ \\
$\lambda$  ($^{\circ}$) & $5.3594249045$ & $197.73278953$ & $4.5221318570$ \\ \hline
$m$ (kg) & $5.997\times10^{10}$ & $3.754\times10^{19}$ & $61.760\times10^{19}$ \\
$d$ (g/cm$^3$) & $0.500$ & $1.152$ & $0.956$ \\
\hline                               
\end{tabular}
\end{table}

\begin{table}
\caption {Physical parameters of Saturn.} 
\centering
\label{saturn}
\begin{tabular}{ll}
\hline \hline 
Radius (km)& $60330$ \\
Mass ($\times 10^{26}$kg)& $5.68683765495$ \\
$J_2$ ($\times 10^{-6}$) & $16290.543820$ \\
$J_4$ ($\times 10^{-6}$) & $-936.700366$ \\
$J_6$ ($\times 10^{-6}$) & $86.623065$\\
\hline                               
\end{tabular}
\end{table}

\begin{figure} 
	\includegraphics[width=\columnwidth]{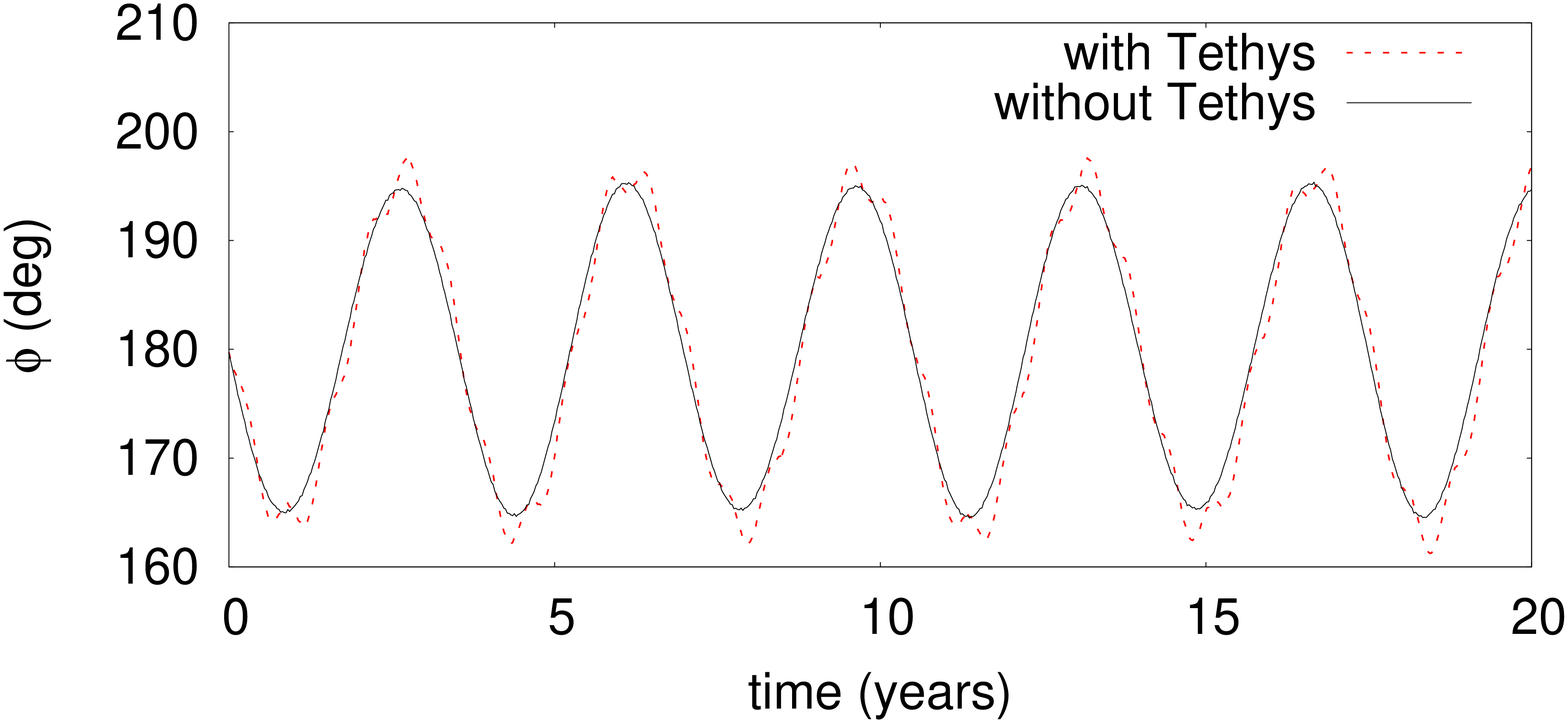}
     \caption{(Colour online) Resonant argument of Aegaeon as a function of time with (dashed line) and without the gravitational effects of Tethys. Tethys disturbs the resonant angle of Aegaeon.}
    \label{cer} 
\end{figure}

The resonant argument of Aegaeon as a function of time is showed in Figure~\ref{cer}. Gravitational effects of the satellite Tethys induce small variations in the resonant argument which can be seen in Figure~\ref{cer} (dashed line). Although these small variations did not alter the lifetime of the particles, the effects of Tethys were added to the system just for completeness.

A sample of 6000 test particles was randomly distributed, with uniform probability, in a resonant arc confined 60~km in radius and  60$^{\circ}$ azimuthally.  The  initial orbital elements, $e, I, \omega$ and $\Omega$,  of  the particles have the same values of the initial  orbital elements of Aegaeon (Table~\ref{sats}). This sample of particles was used in all numerical simulations described in this section. When the distance between the particle and Aegaeon is less than the radius of the small satellite ($r=240$~m, \textcolor{blue}{Hedman et al. 2010}), a collision is detected.

\begin{figure}
\includegraphics[width=\columnwidth]{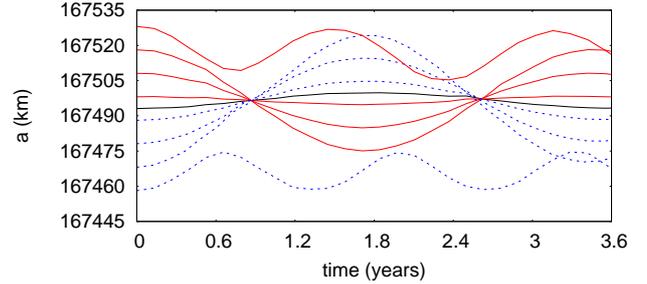}
\caption{(Colour online) This figure shows the geometric semi-major axis (in km) as a function of time (in years) of Aegaeon 
and 8 particles near the outer (full line) and inner (dashed line) edges of the arc.}
\label{smo}
\end{figure}

Our numerical simulations for a timespan of 500~years showed that the particles, due to Mimas 7:6 CER, are azimuthally confined in the arc with an amplitude of  60$^\circ$.  
Figure~\ref{smo} presents the geometric semi-major axis as a function of time of Aegaeon and eight arc particles. These particles are initially displaced by 5, 15, 25 and 35~km from the resonant semi-major axis (167493.73~km). Although the particles displaced by 5-25 km from Aegaeon's semi-major axis show some variation in the semi-major axis, they remain located in the arc region. 
The 7:6 CER with Mimas has a width of about 60~km \citep{elmou14}, therefore those particles displaced by more than 30~km from the resonant semi-major axis are not azimuthally confined.

Besides the gravitational effects of Mimas on the arc particles, the small satellite Aegaeon also causes small variations in their orbital elements. 
Variations in the geometric eccentricity and inclination of the particles are of order $10^{-6}$, and these effects are not strong enough to 
remove these particles from the resonance. Our numerical simulations showed that about 75\% of the initial set of arc particles, under the gravitational effects of 
Saturn, Mimas and Aegaeon, collide with Aegaeon in 500~years. Thus, Aegaeon acts as a sink for these arc particles.

The Cassini cameras observed a population of $\mu$m sized particles located in the G~ring arc. These tiny particles, 1-10$\mu$m in radius ($r$), can be strongly influenced by the effects of the solar 
radiation force. Considering that the planet has its heliocentric position vector as $\mathbfit{r}_{sp}$ ($r_{sp} = |\mathbfit{r}_{sp}|$) and velocity as $\textbf{V}_{P}$, the solar radiation force (SRF) experienced by a circumplanetary particle is \citep{Mi84}

\begin{equation}
\mathbfit{F} = \frac{\Phi A}{c} Q_{pr} \left\lbrace
\left[ 1 - \frac{\mathbfit{r}_{sp}}{r_{sp}}\cdot\left( \frac{\textbf{V}_{P}}{c} + \frac{\mathbfit{V}}{c} \right) \right] \frac{\mathbfit{r}_{sp}}{r_{sp}}
- \frac{\mathbfit{V}_{P} + \mathbfit{V}}{c}
\right\rbrace
\label{E-force}
\end{equation}

\noindent where $c$ is the speed of light and $\mathbfit{V}$ is the velocity vector of the particle with respect to the planet. The particles' cross section is $A$, and we considered that they are made of ideal material which implies $Q_{pr}=1$. In our model the planet is in a circular orbit, hence $r_{sp}$, the magnitude of $\textbf{V}_P$ and the solar flux $\Phi$ are constants. Furthermore, we assumed that the Sun lies in the equatorial plane of the planet (i.e. the obliquity of the planet was neglected), which is a simplification that does not change significantly the magnitude of the solar radiation force 
and saves computational time. We also disregarded secondary and, at least an order of magnitude, weaker effects such as the planetary light reflection and shadow, and the Yarkovsky effect \citep{Ha96}.

We do not include the effects of the plasma drag and the electromagnetic force. 
These forces are responsible to cause an outward drift of the particle and the precession of orbital pericenter, respectively \citep{Su15,Bu01}. 
As discussed in more detail in section~\ref{discussion}, including these forces will probably further reduce the lifetimes of these particles. 

The solar radiation force equation was decomposed and its components were included in the Mercury package \citep{Sf09} in order to analyze the orbital evolution of particles with sizes of 1, 3, 5 and 10$\mu$m in radius. These particles are also perturbed by the gravity of Mimas, Tethys and Aegaeon, and the gravity coefficients of Saturn.

\begin{figure}
\includegraphics[width=\columnwidth]{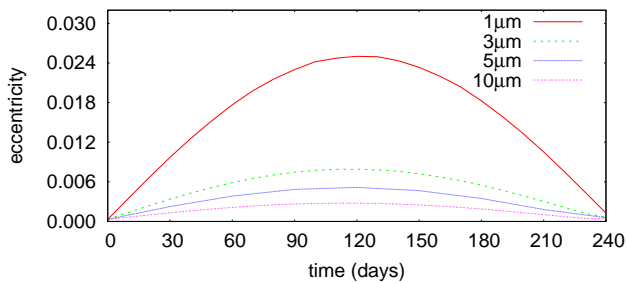}
\caption{(Colour online) Time evolution of the geometric eccentricity of particles with different sizes due to the radiation pressure component. As the size of the particle increases, its $\Delta e$ decreases.}
\label{ecc}
\end{figure}

The radiation pressure component (RP, the term that does not depend on the velocities in Eq.~(\ref{E-force})) mainly causes a variation in the eccentricity
of the particles which can be seen in Figure~\ref{ecc}. Each curve shows the time variation of the eccentricity of four particles, 
initially located at 10~km from the CER semi-major axis, with sizes of 1, 3, 5 and 10~$\mu$m in radius. 
The smaller particle (1~$\mu$m) has the larger variation in the eccentricity, from 0 to $10^{-2}$. The 10~$\mu$m sized particle has the smaller variation in the eccentricity, from 0 to $10^{-3}$.

\begin{figure}
\subfigure[]{\includegraphics[width=\columnwidth]{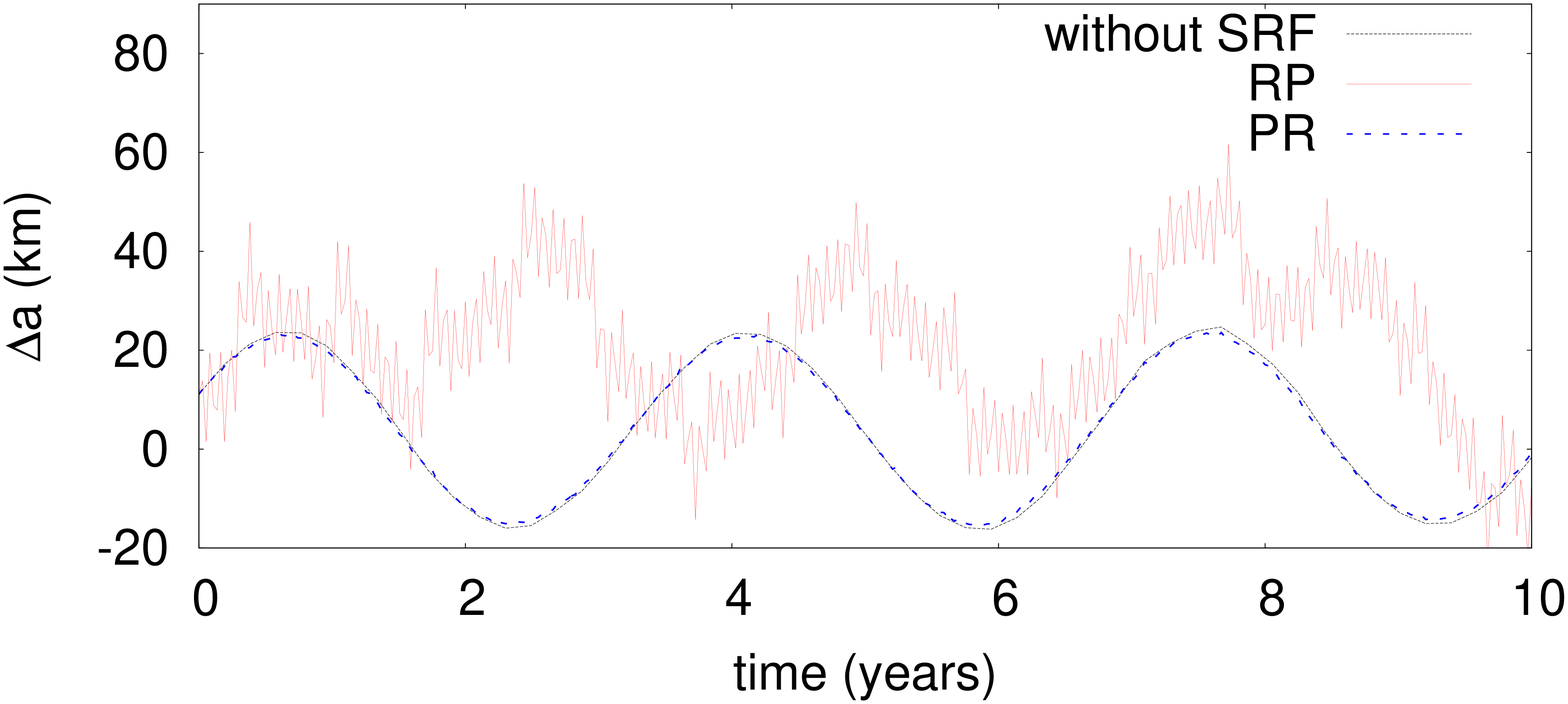}}
\subfigure[]{\includegraphics[width=\columnwidth]{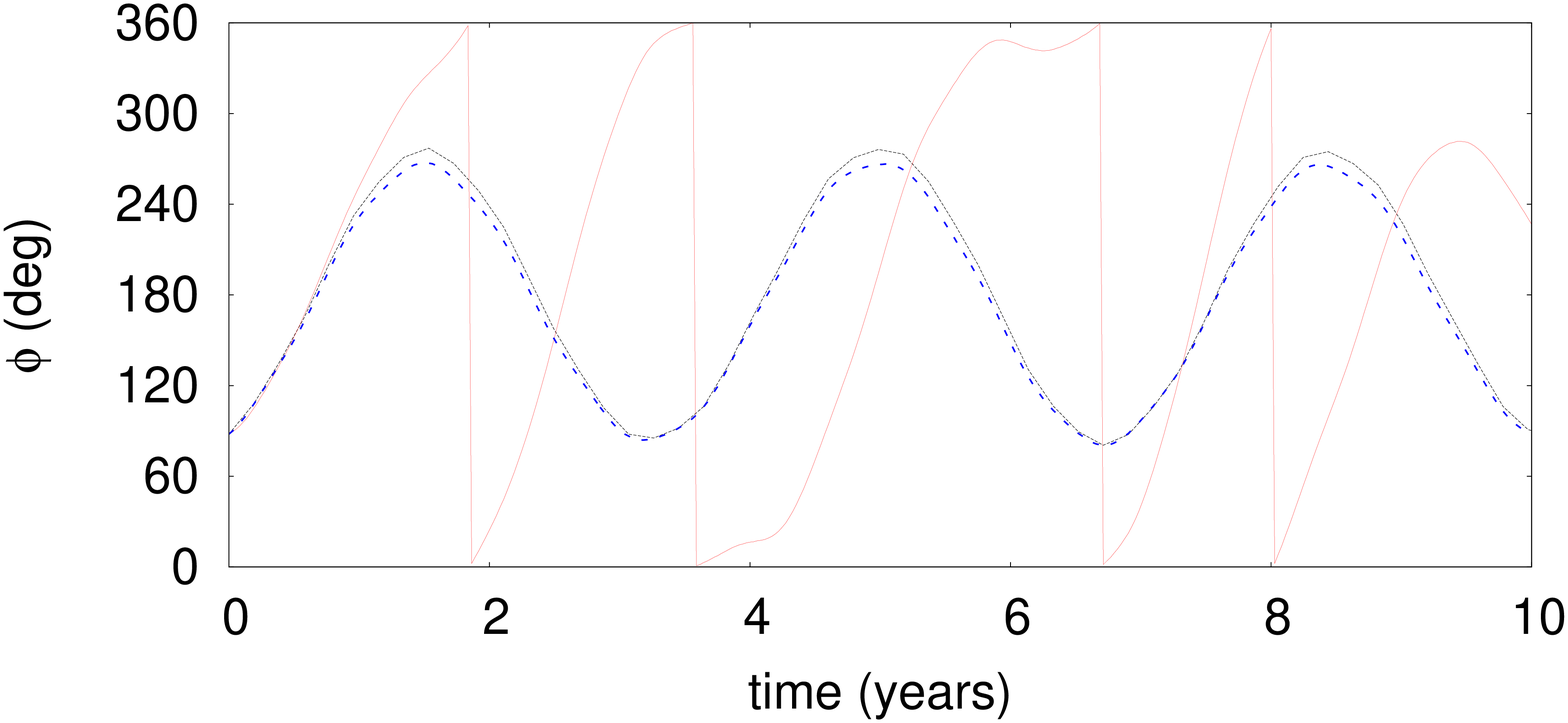}}
\caption{(Colour online) Time variation of (a) the difference between the semi-major axis of the particle and Aegaeon and (b) the resonant argument without the solar radiation force (SRF, dotted line), with the Poynting-Robertson component (dashed line) and with the radiation pressure (full line). The $1 \mu$m sized particle is initially with $\Delta a$=15km and has the same mean anomaly of Aegaeon.}
\label{srf}
\end{figure}

The radiation pressure component provokes, besides a variation in the eccentricity, short-period oscillations (about 40 days) 
in the semi-major axis of the particles, while the Poynting-Robertson component (PR, those terms velocity-dependent in Eq.~\ref{E-force}) 
causes a slow decay of the semi-major axis, but in a timescale much longer than the effects of the RP component. 
Figure~\ref{srf} shows the variation of the semi-major axis ($\Delta a$), and the resonant argument ($\phi$) for a  
1~$\mu m$ in radius particle initially with $\Delta a$=15km, without the SRF and when considering each component separately. 
We can see that the effects of the Poynting-Robertson are negligible in this timescale, while the radiation pressure component causes kilometer variations in the semi-major axis. For this particle, the variation in $a$ is enough to remove the particle from the resonance in less than 2 years.


\begin{figure}
\subfigure[]{\includegraphics[width=\columnwidth]{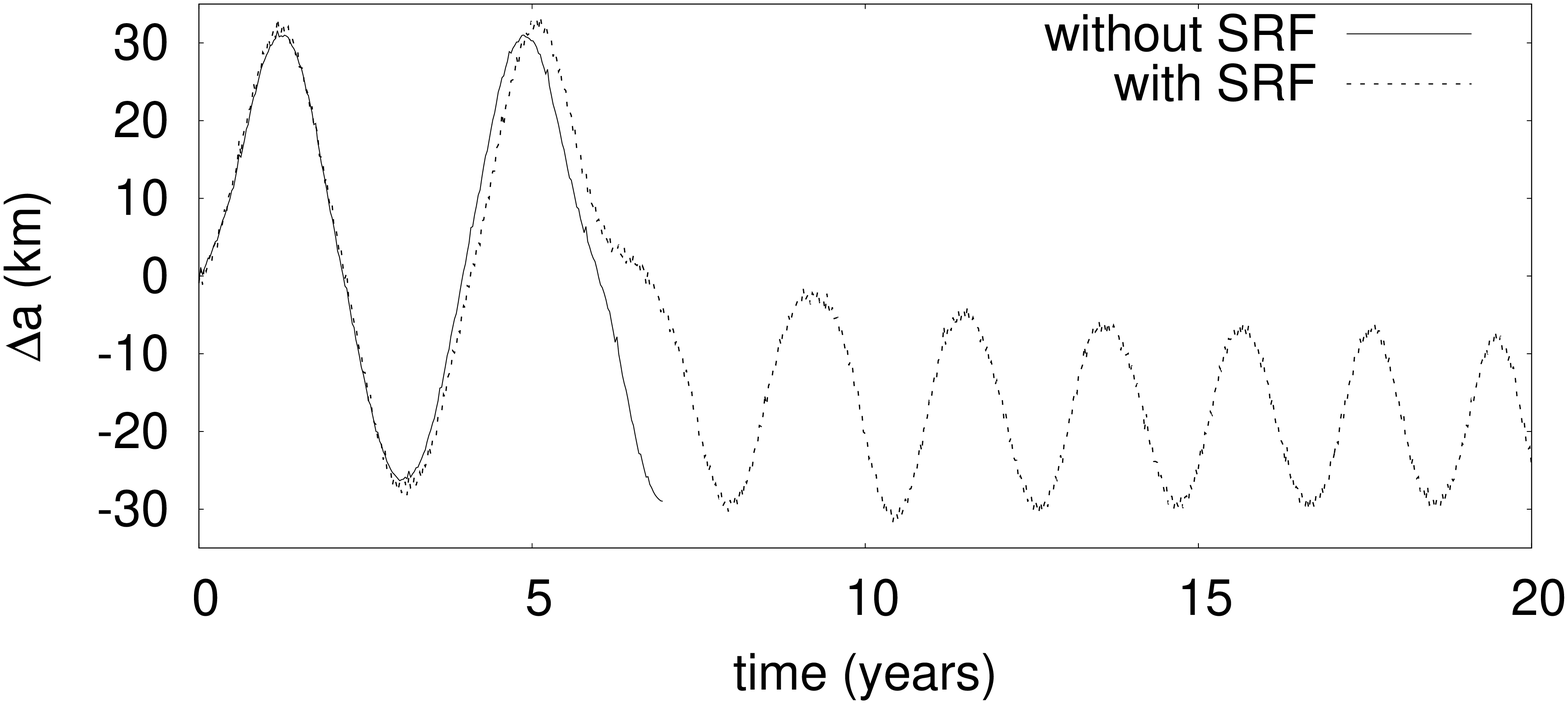}}
\subfigure[]{\includegraphics[width=\columnwidth]{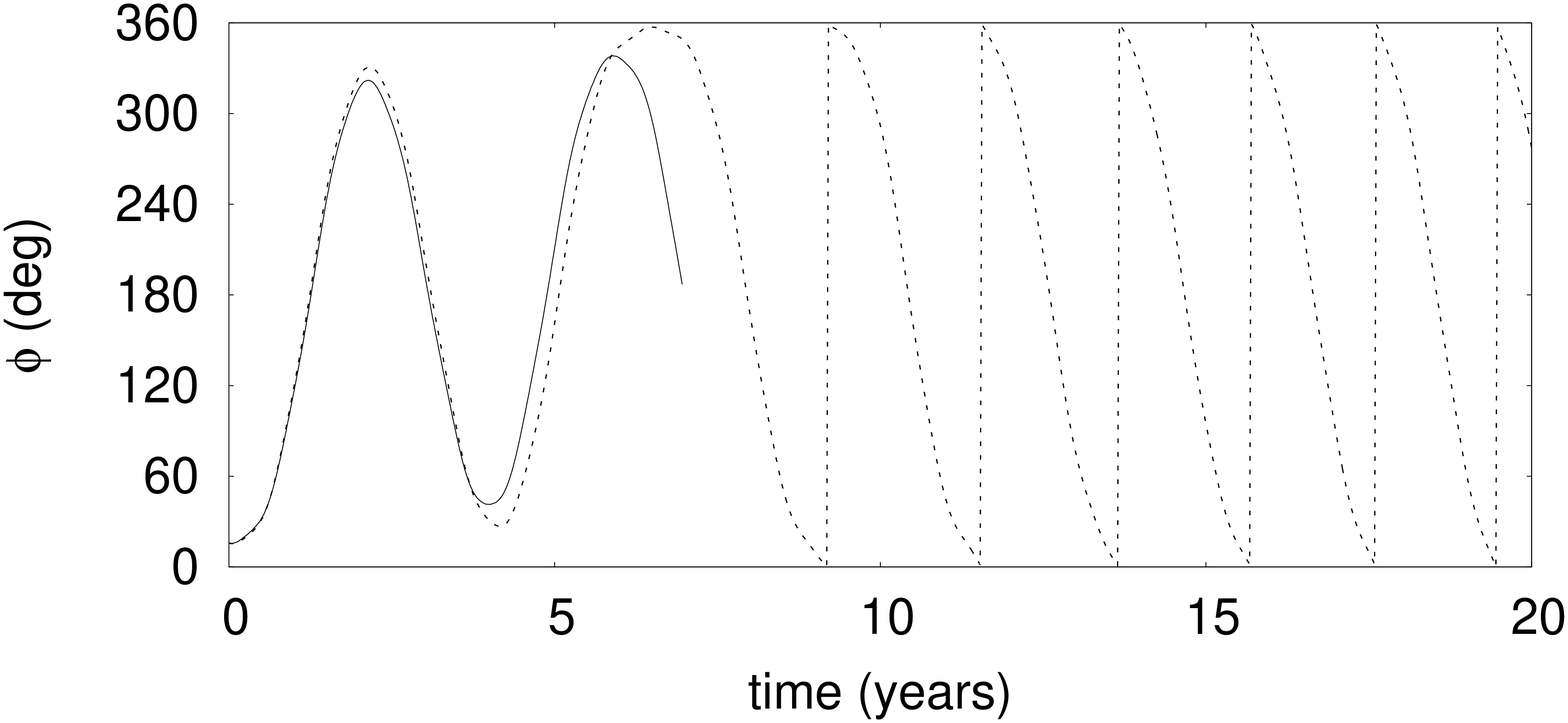}}
\caption{Time variation of (a) the difference between the semi-major axis of a $10\mu$m sized particle and Aegaeon, and (b) the resonant argument with (dashed line) and without the SRF. In both cases the particle is located at the same semi-major axis of Aegaeon and displaced by $\Delta\lambda_0=20^{\circ}$.}
\label{sfrcer}
\end{figure}

Whether or not the particle remains in resonance may change its lifetime. Figure~\ref{sfrcer} shows the difference between the semi-major axis of the particle and Aegaeon, and the resonant argument as a function of time for a $10\mu$m sized particle, initially with the same semi-major axis ($\Delta a_0=0$km) of Aegaeon and mean anomaly displacement by $\Delta\lambda_0=20^{\circ}$, with (dashed line) and without the effects of the solar radiation force. When no solar radiation force is acting in the system, the particle is trapped in the 7:6 CER with Mimas until colliding with Aegaeon in less than 10~years. The effects of the solar radiation force remove the particle from the resonance after about 10~years, when the resonant argument starts to circulate, and  the lifetime of the particle increases by a factor 2.

\begin{figure}
\subfigure[]{\includegraphics[width=\columnwidth]{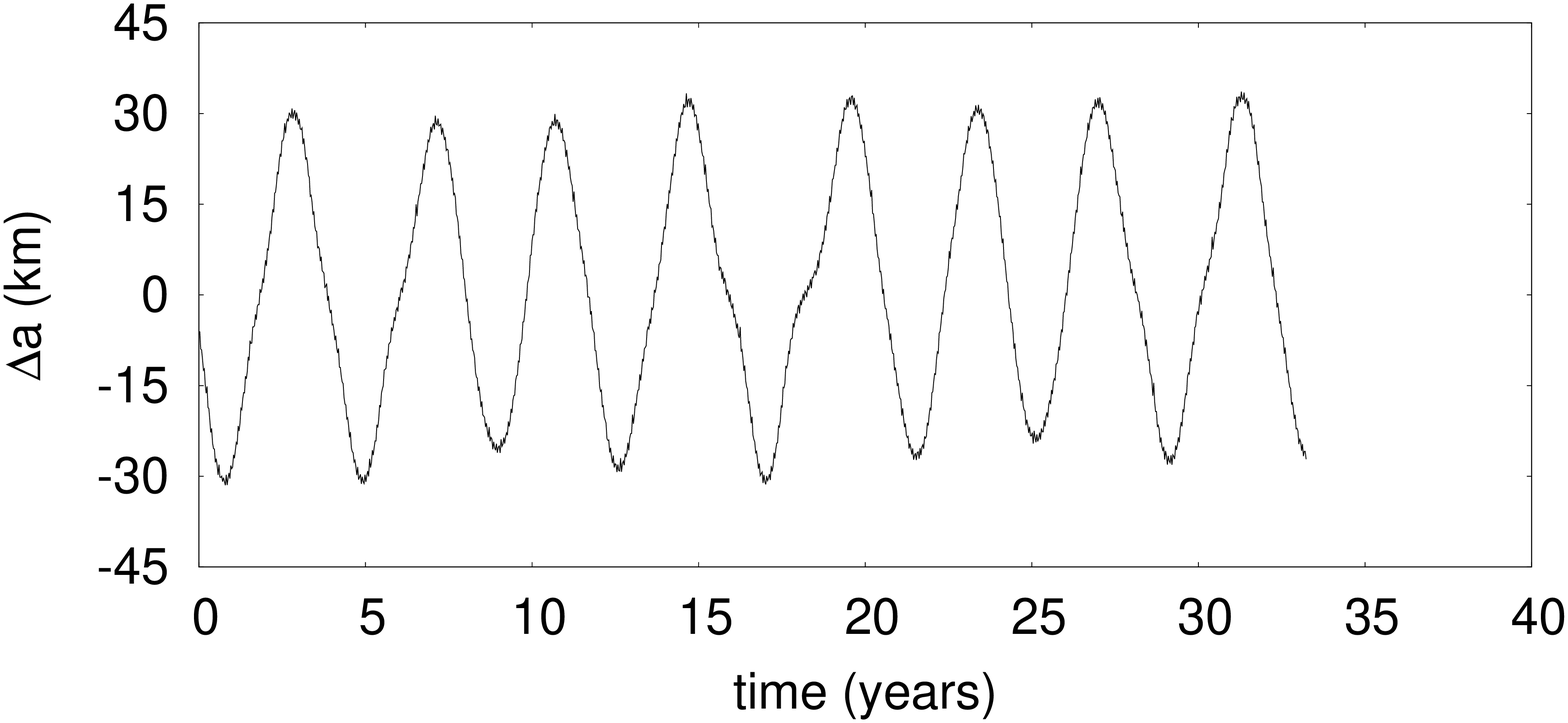}}
\subfigure[]{\includegraphics[width=\columnwidth]{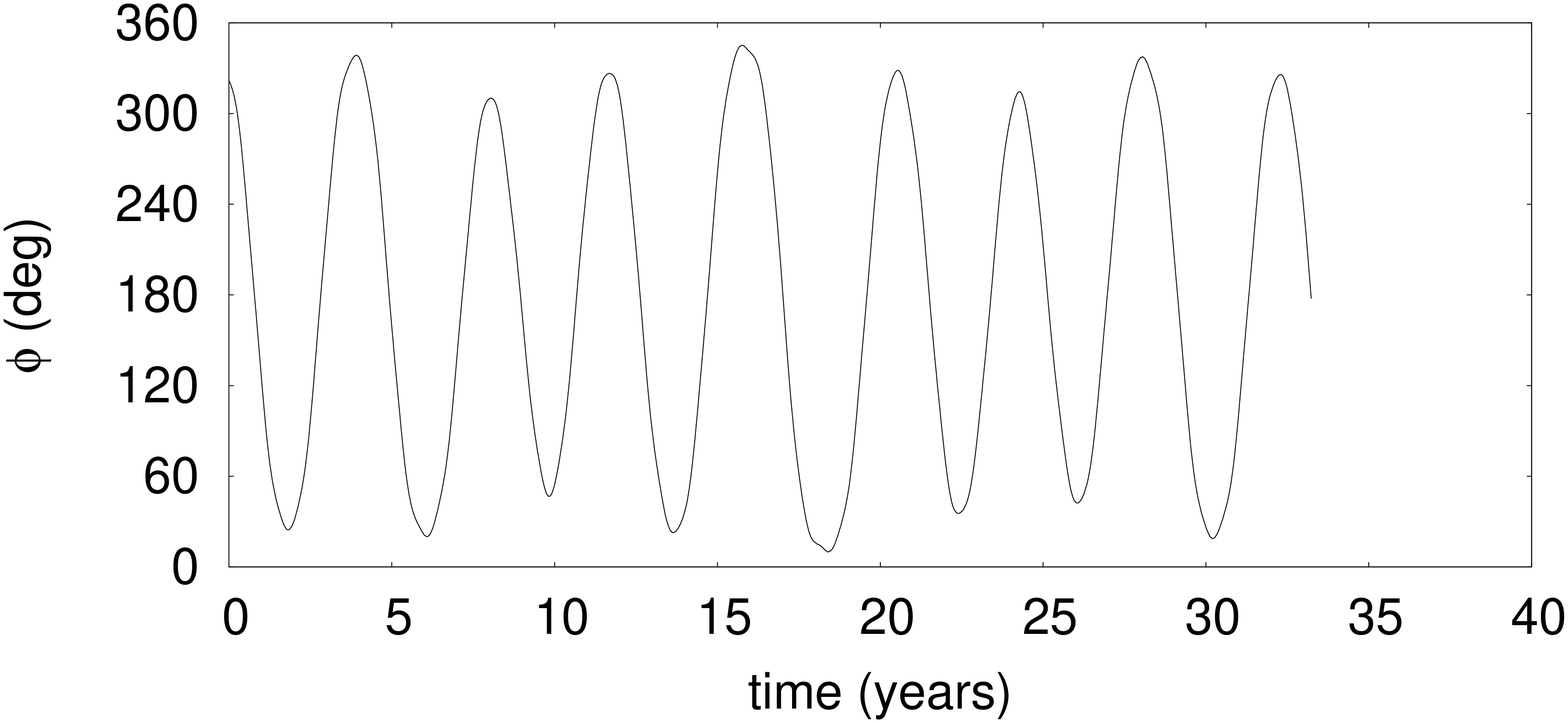}}
    \caption{(a) The difference between the geometric semi-major axis of the particle and Aegaeon and (b) the resonant argument as a function of time in years for a  $10~\mu$m sized particle initially at $\Delta a_0=-5$km and $\Delta\lambda_0=0^{\circ}$. The particle collides with the satellite in less than 35~years.}
    \label{partreso}
\end{figure}

In some cases a particle remains trapped in the resonance, despite of the perturbation of the SRF. Figure~\ref{partreso} shows the difference between the geometric semi-major axis of the particle and Aegaeon and the resonant argument of a representative particle with $10\mu$m in radius, initially with $\Delta a_0=-5$km (a=167488km) and $\Delta\lambda_0=0^{\circ}$ from Aegaeon. This particle remains trapped in the 7:6 CER with Mimas for almost 35~years, until colliding with the satellite.

\begin{figure}    
\subfigure[]{\includegraphics[width=\columnwidth]{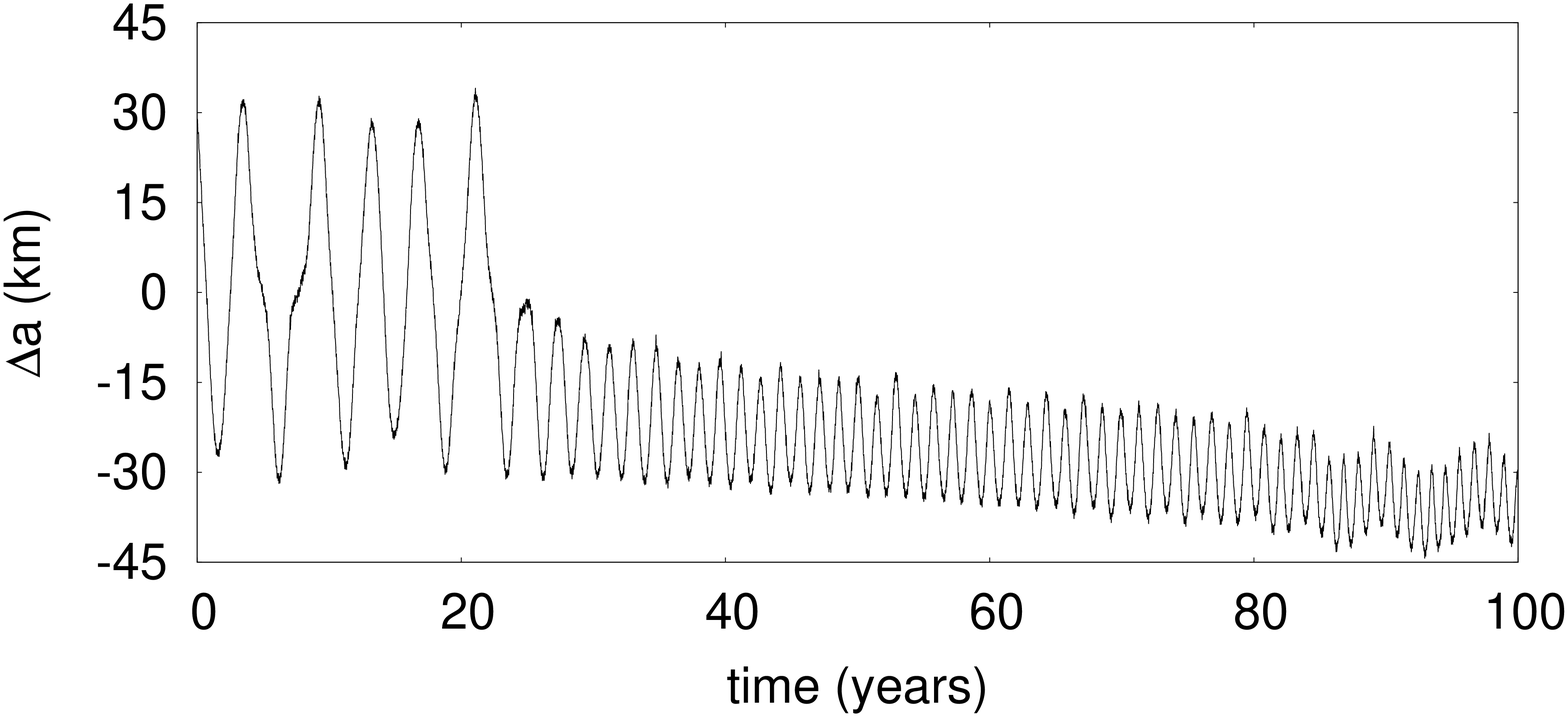}}
\subfigure[]{\includegraphics[width=\columnwidth]{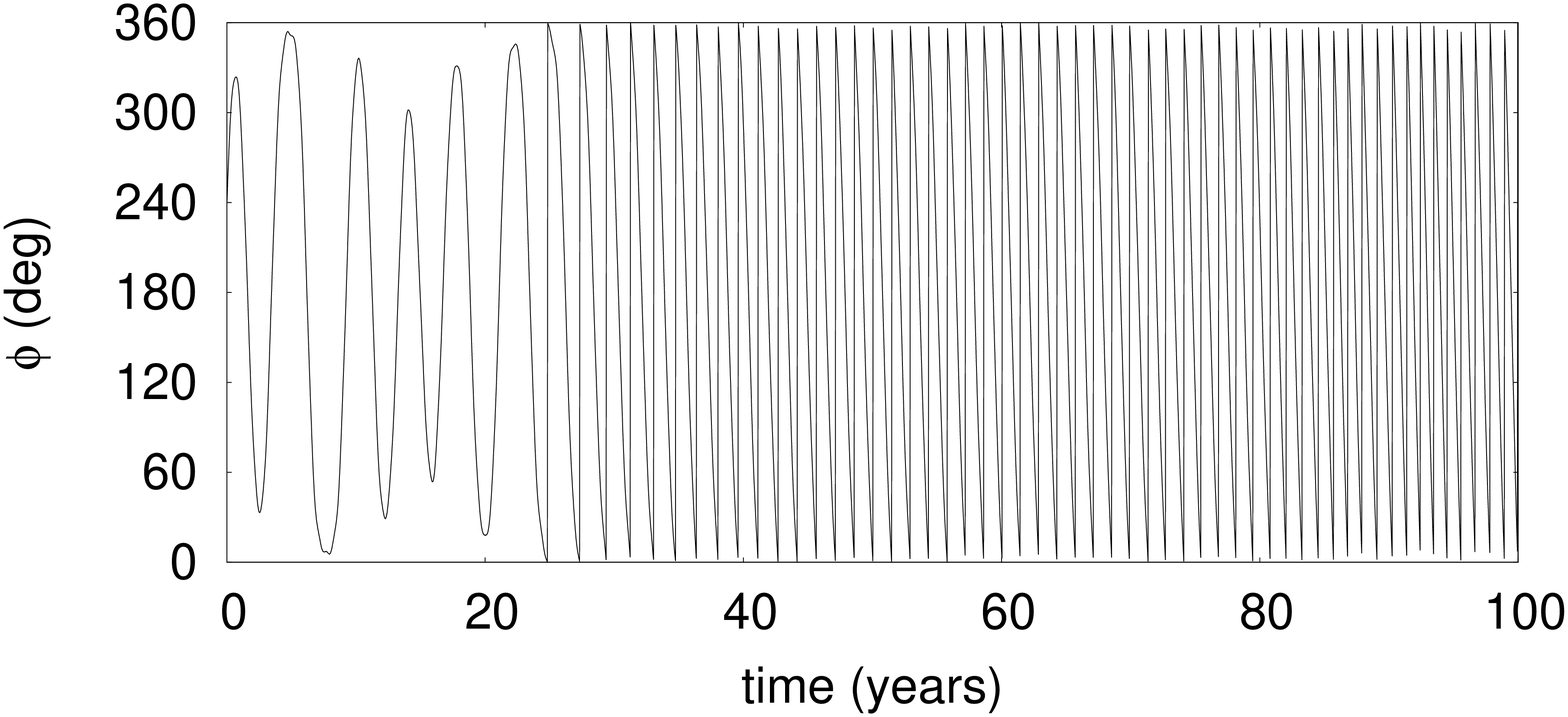}}
   \caption{The same as Figure~\ref{partreso} for a particle initially with $\Delta a_0=~25$~km (a=167522km) and $\Delta\lambda_0=0^{\circ}$.} 
    \label{partnreso}
\end{figure}

We also found some particles that leave the arc, as it is shown in Figure~\ref{partnreso} for a $10~\mu$m grain in radius, initially with $\Delta a_0=25$~km (a=167522~km) and $\Delta\lambda_0=0^{\circ}$. This particle leaves the 7:6 CER with Mimas after about 20~years. After this, the semi-major axis starts to decrease due to the Poynting-Robertson component, and the particle moves away from the arc region to the inner edge of the G~ring.

We then computed the percentage of particles which remain in resonance, and those that the resonance argument 
circulates. The first corresponds to the column `arc' in Table~\ref{azimu}, while the column `ring' corresponds to the percentage of particles that leave the CER. 
It is also presented the time necessary to remove through collisions 90\% of the ``arc" particles ($t_{90}$).

\begin{table}
\caption{Percentage of particles of different sizes that remain in the arc (in resonance) and those particles that leave the arc but remain in  the G~ring. The time (in years) necessary to 90\% of the total ensemble of particles, trapped in the arc, to be removed by collisions is shown in the last column.}
\label{azimu}
\centering
\begin{tabular}{l|cc|c} \hline \hline
$r$ ($\mu$m) &  arc (\%) &   ring (\%)  & $t_{90}$ (years) \\ \hline
1            & 55   & 45   &  3\\
3            & 58   & 42  &  9\\
5            & 60   & 40   &  16\\
10           & 64   & 36   &  26\\
\hline
\end{tabular}
\end{table}

Even though Aegaeon's and the particles' semi-major axes periodically cross due to the resonance, the excitation of the eccentricity caused by the SRF increases the number of times that the particles' orbital radius intersect the orbit of the satellite. Since the amplitude of the eccentricity varies inversely with the particles' size, it is expected that the smaller grains have a larger probability to leave the arc and also have a shorter lifetime. After 3 years, more than 90\% of the 1~$\mu m$ particles collide with the satellite, while 10$\mu$m sized grains have longer lifetimes.

\section{Particles ejected from Aegaeon's surface} \label{ejected}

Throughout the Solar System there is a flux of interplanetary dust particles (IDPs), which can be focused by the presence of a planet. While moving towards the planet, these IDPs can collide with a satellite at a speed of $\mathcal{O}$(10 km/s), and if the satellite is small (radius up to a few tens of kilometers), the outcome of these hypervelocity impacts is the ejection of micrometric particles. Before calculating the amount of material produced by these collisions on Aegaeon surface (section \ref{mass}) we will analyze the orbital evolution of a set of particles ejected from its surface.  

We considered particles with sizes of 1, 3, 5 and 10$\mu$m in radius, leaving the surface of the satellite with initial velocities equal to 1, 5 and 10 times the escape velocity of the satellite ($v_{esc}$). These parameters were chosen in a manner to cover the most likely distribution expected for the dust production mechanism \citep{Kr03}. Similarly to the previous section, the dust grains evolved under the influence of the planet and its gravity coefficients ($J_2$, $J_4$ and $J_6$), the gravitational effects of Mimas, Tethys and Aegaeon, and also the solar radiation force. For each combination of particle size and ejection velocity, an ensemble of 1,000 particles, launched at the same time, was analyzed. All particles were launched radially away from Aegaeon and in the equatorial plane of Saturn. The angular position, related to the surface of the satellite, was randomly chosen from 0 to $360^{\circ}$ with uniform probability.

Table \ref{T-ejecao} summarizes the outcome of the numerical simulations. We divided each sample into \textit{arc} particles, corresponding to those remaining all the integration time in resonance (thus azimuthally confined), and the \textit{ring} particles the ones that make excursions through the ring. Since all particles hit the surface of Aegaeon we computed the time necessary to remove $90\%$ of each set ($t_{90}$).

\begin{table}
	\caption{Summary of the numerical simulations considering particles of different sizes and ejection velocities. The particles were classified as \textit{arc} when they remain the entire simulation in resonance and as \textit{ring} otherwise. $t_{90}$ corresponds to the time in years necessary to 90\% of the ensemble to be removed by collision.}
\centering
\begin{tabular}[!hb]{ccllll} 
		\hline \hline
         &               & 1$\mu$m & 3$\mu$m      & 5$\mu$m& 10$\mu$m                    \\ \hline
\multirow{4}{*}{$1v_{esc}$}    & arc           & 100\%    & 100\%               & 100\%               & 100\%               \\
       & $t_{90}$    & 2.5                  & 4.6                 & 14.4                 & 38.2                 \\ 
       & ring          & 0\%                 & 0\% & 0\%                 & 0\%                 \\
       & $t_{90}$ & - & - & - & - \\ \hline
\multirow{4}{*}{$5v_{esc}$} & arc           & 92\% & 100\%               & 100\%               & 100\%               \\
           & $t_{90}$  & 3.0 & 12.6                 & 20.6                 & 36.7                \\
           & ring          & 8\%  & 0\%                 & 0\%                 & 0\%                 \\
           & $t_{90}$ & 7.1 & - & - & - \\ \hline
\multirow{4}{*}{$10v_{esc}$} & arc           & 66.3\% & 68.7\%                & 70.6\%                & 71.3\%                 \\
            & $t_{90}$  & 3.4 & 13.9                 & 22.5                 & 41.9                \\
            & ring          & 33.7\% & 31.3\%                & 29.4\%                & 28.7\%                \\
            & $t_{90}$  & 35.0 & 94.9                & 136.3               & 278.6 \\ \hline
            \hline
	\end{tabular}
\label{T-ejecao}
\end{table}

The amount of particles that remain in the arc decreases slightly for smaller grains, but we note a substantial change when the ejection velocity increases. 
Smaller and faster grains are more likely to leave the arc and therefore they can survive longer. It is mainly caused by those particles that experience stronger variations in the semi-major axis due to the SRF and, as a consequence, they leave the resonance. At this point those particles are classified as `ring' and they present larger survival time, since their orbital paths go further from the satellite.

When launched at lower speeds, the particles remain entirely confined in the arc. Their survival time is shorter since they stay closer to Aegaeon's orbit. The longer lived ensemble is the one formed by the faster and larger grains, but even in this case the particles do not survive more than 300~years.

\section{Mass Production Rate} \label{mass}
Besides acting as a sink for the particles of the G~ring/arc, Aegaeon can produce dust due to the impacts of interplanetary grains. Through the simplified model presented in \cite{Sf12} we compute the mass production rate ($M^+$) due to the impacts of projectiles directly onto the surface of the satellite as
\begin{align}
M^+  = F_\imp Y S
\label{E-mplus}
\end{align}
\noindent  where $F_\imp$ is the mass flux of impactors that reaches the satellite, $Y$ is the ejecta yield and $S$ is the satellite cross section.

The characterization of the interplanetary dust grain environment is a difficult task, specially for the outer part of the Solar System due to the small number of direct measurements. Combining data from several missions, \cite{Po16} estimates that the mass flux at Saturn's heliocentric distance is $1 \times 10^{-17}$~kg/(m$^2\cdot$s). This value is enhanced by the gravitational focusing due to the planet, 
so that the effective flux at Aegaeon's orbit is $F_\imp \sim 5.5 \times 10^{-17}$~kg/m$^2$/s. It was assumed that the radius of Aegaeon is 240~m, so its cross section area is 
{$\sim 1.8\times 10^5$ m$^2$}.

The yield measures efficiency of the ejection process, and for a satellite with a pure ice surface (no silicates) it can be written as \citep{Ko11}
\begin{align}
Y = 2.64 \times 10^{-5}  m_{\rm imp}^{0.23}~v_{\rm imp}^{2.46}. 
\end{align}
For a typical impactor of $10^{-8}$~kg  ($r\sim 100\mu$m) with velocity of $23$~km/s (after the gravitational focusing), 
the yield  is $Y \sim 2 \times 10^{4}$.

By this mechanism, impacts with Aegaeon produce dust grains at a rate of
\begin{align}
M^+ \sim 2 \times 10^{-7}~\textrm{kg/s}.
\end{align} 

In order to determine if ejecta from Aegaeon can be the source of visible dust in the arc, 
it is  necessary to estimate the mass of the arc. First we assume a power law distribution for the particle size distribution as
\begin{align}
dN = Cr^{-q}~dr
\label{Eu-dn}
\end{align}
\noindent where $dN$ is the number of particles and $q=3.5$ is assumed as a typical value. The optical depth can be written as
\begin{align}
\tau = \int_{r_1}^{r_2} d\tau = \int_{r_1}^{r_2} \pi r^2 ~dN
\label{Eu-tau-op}
\end{align}
\noindent and the mass of the dust can be calculated by \citep{Sf12}
\begin{align}
m =& A_{\textrm{arc}} \left(\frac{4}{3}\pi  \rho\right) \int_{r_1}^{r_2}   r^3 ~dN 
\label{Eu-massa}
\end{align}
A$_{arc}$ is the surface area of the arc. We assumed a simplified model considering the arc as a $60^{\circ}$ circular sector with radius $167493 \pm 125$km. 
If we consider an uniform optical depth for the arc as $\tau=10^{-5}$ \citep{He07} and dominated by small ice particles ($r=[1-10]\mu$m), which is expected by 
the impact process, it gives $m \sim 2\times10^6$~kg. This value is close to the high end of the dust mass estimates from \cite{He07}.

Neglecting any loss mechanism, the total amount of dust in the arc would need about 30,000 years to accumulate. Since this time is at least 
three orders of magnitude larger than those presented in Table~\ref{T-ejecao}, it is unlikely that Aegaeon alone could keep the arc dust in a steady state.

\section{Discussion} \label{discussion}
The G ring arc region is a dynamic environment composed by dust particles, probably cm-m sized bodies, and a small satellite Aegaeon. The satellite and the particles are both trapped in a 7:6 CER with the satellite Mimas, responsible for keeping the particles and Aegaeon azimuthally confined in $60^{\circ}$ of longitude. Besides the gravitational effects, the $\mu$m sized dust particles are also strongly affected by the solar radiation force, which can lead them to collisions or ejection from the arc.

In this work we analyze Aegaeon's effects on the G-ring arc. In the numerical simulations the set of $\mu$m sized particles is under the gravitational effects of the massive bodies Saturn, Mimas, Tethys and Aegaeon. Our results showed that Aegaeon acts as a sink for the particles, removing them by collisions. About 75\% of the confined particles are removed from the arc in less than 500~years.\\
\indent The solar radiation component induces short period variations in the semi-major axis of the particles, which changes the resonant argument, and as a result most of the particles are removed from the resonance. This force also changes the eccentricities of the particles, leading to orbital crossing and eventually collisions between these particles and the satellite Aegaeon. The lifetime of 90\% of the initial set of 1~$\mu$m sized particles is about 3~years, and 26~years for  particles 10~$\mu$m in radius. Therefore, the presence of Aegaeon reduces the lifetime of the particles leading to the extinction of the arc. This lifetime may be even shorter, since in our numerical simulations the plasma drag and the electromagnetic force were not taken into account. The former perturbation causes an outward drift \citep{Su15}, while the latter may increase even more the eccentricity of the particles \citep{Ha93}, thus both perturbations will probably cause particles to be removed from the arc even faster.\\
\indent \cite{He09} argued that debris ejected from the surface of small satellites can stay confined in the same resonance of these satellites. Our results showed it is true 
for the particles ejected at slower velocities. However, the confined particles have a very short lifetime, 90\% of the entire population leave the arc in less than 20~years. Only 8\% of the smaller particles goes to the G~ring, the majority of them collide with the satellite Aegaeon. The longer lived ensemble is composed by larger particles leaving the surface of the satellite with 10$v_{esc}$, these particles can last up to $\sim300$~years.\\
\indent Although Aegaeon is below the optimum size to generate dust by impacts of interplanetary dust particles \citep{Bu01} we calculated the amount of dust 
this small satellite can contribute to the arc population. Our result shows that Aegaeon can produce particles at a rate of $2 \times 10^{-7}$~kg/s, and if the mass of the arc is about $2 \times 10^6$~kg, neglecting any loss mechanism, it would take more than 300,000~years to Aegaeon populated the arc. Our simulations therefore show that Aegaeon is probably a net sink for arc particles.\\
\indent Our simplified model takes into account only the direct process of dust. The assumed yield and interplanetary dust flux must be taken at least with one order of uncertainty. However, even considering a lower estimate for the arc mass and the upper limits for the flux and yield, the time necessary to accumulate such amount of dust is several orders of magnitude larger than the survival time of the particles.\\
\indent In order to maintain the dust population in a steady state additional processes must be invoked. For instance, there is evidence that the ring may be populated by multiple objects $\mathcal{O}$(cm-m) across, which are below the threshold level of the cameras to be detected. A more complex process involving secondary impacts of IDPs with these objects, or even impacts among themselves, could produce more dust particles than the primary impacts.

\section{Acknowledgements}
We would like to thank the anonymous reviewer who greatly improved the final text. The authors thank Fapesp (Proc.~2016/2488-0 and Proc.~2011/08171-3) and CNPq (Proc.~309714/2016-8 and Proc.~305737/2015-5) for the financial support.







\bsp	
\label{lastpage}
\end{document}